\documentclass[twocolumn,showpacs,amsmath,amssymb]{revtex4}
\usepackage{graphicx}
\usepackage{color}

\begin{document}

\title{Exact-Diagonalization Analysis of Composite Excitations in the $t$-$J$ Model}
\author{Takashi Otaki, Yuta Yahagi, and Hiroaki Matsueda\thanks{matsueda@sendai-nct.ac.jp}}
\affiliation{
Sendai National College of Technology, Sendai 989-3128, Japan
}
\date{\today}
\begin{abstract}
We examine spectral properties of doped holes dressed with surrounding spin cloud in the $t$-$J$ model. These composite-hole excitations well characterize prominent band structures in the angle-resolved photoemission spectrum. In one-dimensional (1D) case at half-filling, we identify the composite operators that separately pick up the spinon and holon branches, respectively. After hole doping, we find that the composite hole excitations with string-like spins tend to be localized at $k=\pi/2$ in the momentum space. This means that such composite excitations should be actual electronic excitations, since the spinon and holon branches merge together at this momentum. In 2D case, we find that the composite excitations with more non-local spin fluctuation have stronger intensity near the Fermi level. The composite band structure along diagonal $(0,0)$-$(\pi,\pi)$ direction in 2D has some similarity to that in 1D, and such non-local spin fluctuation plays an important role on the formation of the pseudogap in high-$T_{c}$ cuprates.
\end{abstract}
\pacs{74.25.Jb, 71.10.Fd, 74.72.Kf, 74.25.Ha}
\maketitle

\section{Introduction}

The $t$-$J$ model is one of prototypical examples for understanding strongly correlated electron systems. The interplay between spin and charge degrees of freedom is a key ingredient of this model. In spatially one-dimensional (1D) case, this model (or equivalently the large-$U$ Hubbard model) has been fully understood by the exact and numerical approaches~\cite{Ogata,Penc}. Therein, the electronic band dispersion splits into the spinon and holon branches, and this is called the spin-charge separation. They are collective excitations, not the quasiparticles. In 2D case, it is believed that this model captures anomalous features of hole-doped high-$T_{c}$ superconductivity in cuprates~\cite{Maekawa}. Although it is difficult to solve it exactly, various types of numerical studies are proceeding. The value of this model as the prototypical and minimal correlated system is still not lost, even though roughly 30 years have passed already after the discovery of high-$T_{c}$ cuprates.

Generally speaking, the spin and charge degrees of freedom are separated in 1D, while they are interacting with each other in 2D. However, this statement is sometimes misleading, and for example in the context of the slave boson theory the phase string connects the spinon with the holon excitation~\cite{Suzuura}. The notion of the spin-charge separation and coupling may depend on what we are going to observe and on how to treat the spin and charge degrees of freedom. In this paper, we would like to revisit the coupling of spin and charge degrees of freedom in 1D.

Mathematically, the fundamental operator of the $t$-$J$ model corresponds to a fermionic particle with no double occupation at each site. This operator does not obey the exact fermionic anti-commutation relation. Owing to this constraint, the equation of motion for this excitation necessarily induces more non-local spin excitations~\cite{Matsumoto}, and the equations of motion do not close. A doped hole into the Mott insulator becomes a composite particle dressed with the spin excitation. This mathematical structure is common to both of 1D and 2D cases, and thus it is natural to consider that this non-locality is related to the phase string in 1D and the spin cloud in 2D. The purpose of this paper is to examine the spectral properties of such composite excitation modes. Based on the comparison between 1D and 2D cases, we would like to also discuss about the anomalous band structure of high-$T_{c}$ cuprates.

By this approach, it is possible to do more advanced spectroscopic calculation rather than the standard angle-resolved photoemission spectroscopy (ARPES). In terms of composite operators, the complicated band structures of correlated systems are composed of various types of composite particles. Different branches are represented by different composite operators, respectively. Even if the single-particle spectrum is heavily damped by the strong self-energy correction at a particular energy-momentum region, we may find the composite excitation that is well defined and has a long life time at the region. In other words, we would like to find the best representation of the collective mode at given energy and momentum.

The organization of this paper is as follows. In the next section, we introduce the technical details. In particular, we explain the physical meaning of composite excitations. In Secs. III and IV, we present numerical results for 1D and 2D cases. Based on these results, we compare 1D with 2D results, and finally we summarize our paper.

\section{Method}

\subsection{Model and Fundamental Algebra}

We start with the $t$-$J$ model. The Hamiltonian is defined by
\begin{eqnarray}
H &=& -\sum_{i,j}t_{ij}\xi_{i}^{\dagger}\xi_{j}+\frac{J}{4}\sum_{<ij>}\vec{n}_{i}\cdot\vec{n}_{j},
\label{H}
\end{eqnarray}
where $\xi_{i}^{\dagger}=(\xi_{i,\uparrow}^{\dagger},\xi_{i,\downarrow}^{\dagger})$ is the spinor representation of singly-occupied particle at site $i$, $\xi_{i,\sigma}=c_{i,\sigma}(1-n_{i,-\sigma})$ with spin $\sigma$, $\xi_{i}^{\dagger}\xi_{j}=\sum_{\sigma}\xi_{i,\sigma}^{\dagger}\xi_{j,\sigma}$, $n_{i}=\xi_{i}^{\dagger}\xi_{i}=n_{i,0}=\xi_{i}^{\dagger}\sigma_{0}\xi_{i}$, and $\vec{n}_{i}=\xi_{i}^{\dagger}\vec{\sigma}\xi_{i}$ with the identity matrix $\sigma_{0}$ and the Pauli matrices $\vec{\sigma}=(\sigma_{1},\sigma_{2},\sigma_{3})$. This model includes electron hopping, $t_{ij}$, and the superexchange interaction $J$. Here, the hopping term between the nearest neighbor pair of sites $i$ and $j$ is represented as
\begin{eqnarray}
t_{ij}=t\alpha_{ij}.
\end{eqnarray}
The Fourier transform of $\alpha_{ij}$ is given by
\begin{eqnarray}
\alpha(k)=2\cos k,
\end{eqnarray}
for the 1D chain, and is also given by
\begin{eqnarray}
\alpha(\vec{k}) = 2\left(\cos k_{x}+\cos k_{y}\right),
\end{eqnarray}
for the 2D square lattice.

It is noted that the excitation $\xi_{i}$ is not a simple fermion, and satisfy the following algebrae
\begin{eqnarray}
&& \bigl\{\xi_{i},\xi_{i}^{\dagger}\bigr\} = 1+\frac{1}{2}\sigma^{\mu}n_{i,\mu} , \\
&& \sigma^{\mu}n_{i,\mu} = -n_{i}+\vec{\sigma}\cdot\vec{n}_{i} , \\
&& \xi_{i}n_{i,\mu} = \sigma_{\mu}\xi_{i} , \\
&& \bigl[ n_{i,\mu} , n_{i,\nu} \bigr] = 2i\epsilon_{\mu\nu\lambda}n_{i,\lambda} ,
\end{eqnarray}
where $1=\sigma_{0}$. As we can easily imagine, the first equality necessarily produces spin and charge fluctuation. As we go to higher order hierarchy of the Heisenberg equation of motion for an excitation operator, the fundamental excitation is dressed with such quantum fluctuation.

\subsection{Hierarchy of Composite Excitations and Their Propagators}

We consider the Heisenberg equation of motion for $\xi_{i}(t)=e^{-iHt}\xi_{i}e^{iHt}$ (we take $\hbar=1$). This equation of motion does not close, and induces spatially more extended composite excitations. The equation for $\xi_{i}(t)$ is given by
\begin{eqnarray}
i\frac{\partial}{\partial t}\xi_{i}(t) &=& -t\left(1+\frac{1}{2}\sigma^{\mu}n_{i,\mu}(t)\right)\xi_{i}^{\alpha}(t) \nonumber \\
&& + \frac{J}{2}\vec{n}_{i}^{\alpha}(t)\cdot\vec{\sigma}\xi_{i}(t),
\end{eqnarray}
where we define $\xi_{i}^{\alpha}=\sum_{j}\alpha_{ij}\xi_{j}$. Then, we find the new excitations $\vec{n}_{i}\cdot\vec{\sigma}\xi_{i}^{\alpha}$ and $\vec{n}_{i}^{\alpha}\cdot\vec{\sigma}\xi_{i}$ in which a doped hole combines with the nearest neighbor spin fluctuation. When we next consider the equation of motion for these composite states, we find $\vec{n}_{i}\cdot\vec{\sigma}\xi_{i}^{\alpha\alpha}$ and $\vec{n}_{i}\cdot\vec{\sigma}\left(\vec{n}_{i}\cdot\vec{\sigma}\xi_{i}^{\alpha}\right)^{\alpha}$. For instance, the important excitations up to $L=5$ for 2D $L\times L$ lattice are
\begin{eqnarray}
\Psi=
\left(
\begin{array}{c}
\Psi_{1} \\
\hline
\Psi_{2} \\
\hline
\Psi_{3}^{-} \\
\Psi_{3}^{+} \\
\hline
\Psi_{4}^{--} \\
\Psi_{4}^{-+} \\
\Psi_{4}^{+-} \\
\Psi_{4}^{++}
\end{array}
\right)=\left(
\begin{array}{c}
\xi \\
\hline
\vec{n}\cdot\vec{\sigma}\xi^{\alpha} \\
\hline
\vec{n}\cdot\vec{\sigma}\xi^{\alpha\alpha} \\
\vec{n}\cdot\vec{\sigma}\bigl(\vec{n}\cdot\vec{\sigma}\xi^{\alpha}\bigr)^{\alpha} \\
\hline
\vec{n}\cdot\vec{\sigma}\xi^{\alpha\alpha\alpha} \\
\vec{n}\cdot\vec{\sigma}\bigl(\vec{n}\cdot\vec{\sigma}\xi^{\bar{\alpha}}\bigr)^{\bar{\alpha}\bar{\alpha}} \\
\vec{n}\cdot\vec{\sigma}\bigl(\vec{n}\cdot\vec{\sigma}\xi^{\bar{\alpha}\bar{\alpha}}\bigr)^{\bar{\alpha}} \\
\vec{n}\cdot\vec{\sigma}\bigl(\vec{n}\cdot\vec{\sigma}\bigl(\vec{n}\cdot\vec{\sigma}\xi^{\alpha}\bigr)^{\alpha}\bigr)^{\alpha}
\end{array}
\right).
\end{eqnarray}
These excitations have large spectral intensities that are defined as the diagonal parts of $\left<0\right|\left\{\Psi,\Psi^{\dagger}\right\}\left|0\right>$~\cite{Matsueda}. Thus if we look at the spectra of these composite excitations, we may say the origin of each prominent band structure in the ARPES spectrum. We have confirmed that the excitation originating from the exchange term, $\Phi_{2}=\vec{n}^{\alpha}\cdot\vec{\sigma}\xi$, show similar behavior to $\Psi_{2}$, and hereafter we do not consider it.

In this paper, we calculate the diagonal part of the following propagator matrix
\begin{eqnarray}
G^{R}(\vec{k},\omega) &=& \left<0\right|\Psi_{\vec{k},\sigma}\frac{1}{\omega+E_{0}-H+i\epsilon}\Psi_{\vec{k},\sigma}^{\dagger}\left|0\right> \nonumber \\
&& + \left<0\right|\Psi_{\vec{k},\sigma}^{\dagger}\frac{1}{\omega-E_{0}+H+i\epsilon}\Psi_{\vec{k},\sigma}\left|0\right> \nonumber \\
\end{eqnarray}
where $\left|0\right>$ is the ground state, and $\Psi_{\vec{k},\sigma}$ is the Fourier component of $\Psi_{i,\sigma}$. The spectra for the composite particles are given by the imaginary part of the retarded propagator matrix
\begin{eqnarray}
A(\vec{k},\omega)=-\lim_{\epsilon\rightarrow 0}\frac{1}{\pi}{\rm Im}G^{R}(\vec{k},\omega).
\end{eqnarray}
For instance, the usual single-particle spectrum is given by $A_{\Psi_{1}\Psi_{1}^{\dagger}}(\vec{k},\omega)$. Hereafter, we particulary focus on $A_{\Psi_{2}\Psi_{2}^{\dagger}}(\vec{k},\omega)$, $A_{\Psi_{3}^{+}\Psi_{3}^{+\dagger}}(\vec{k},\omega)$, and $A_{\Psi_{4}^{++}\Psi_{4}^{++\dagger}}(\vec{k},\omega)$.

\subsection{Numerical Conditions}

In 1D (2D) case, we take $16$ ($4\times 4$) lattice under the periodic boundary condition. In doped cases, we take $2$ holes ($14$ electrons), and thus the hole doping rate is $\delta=0.125$. The Hamiltonian matrix is exactly diagonalized by the Lanczos algorithm for the calculation of the propagators for composite excitations. The propagators are also calculated by the kernel polynomial expansion, and we have confirmed the relevance between these methods. The broadening factor is taken to be $\epsilon=0.2t$. Since the momentum resolution of each spatial direction is $\pi/2$ owing to severe finite-size effect, we naturally interpolate the numerical data that the mesh size becomes $\pi/20$. We should note that this procedure does not produce new fine structures originating with lower energy physics. We just introduce this method so that our data are visible.

\section{Numerical Results I: 1D case}

\subsection{Half-filling}

\begin{figure}[htbp]
\begin{center}
\includegraphics[width=8.5cm]{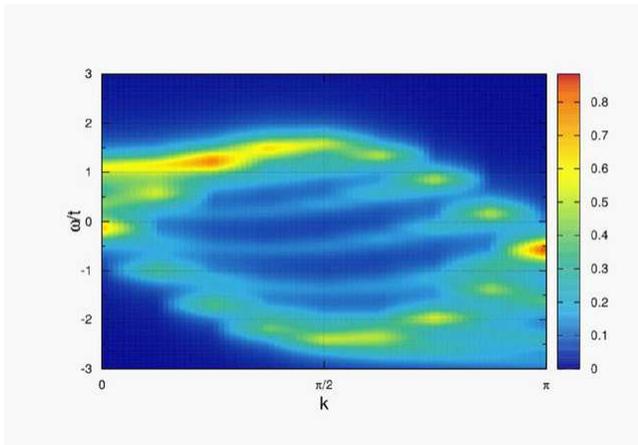}
\end{center}
\caption{ARPES spectrum (spectrum for the bare fermionic particle $\Psi_{1}=\xi$)}
\label{fig1a}
\end{figure}

\begin{figure}[htbp]
\begin{center}
\includegraphics[width=8.5cm]{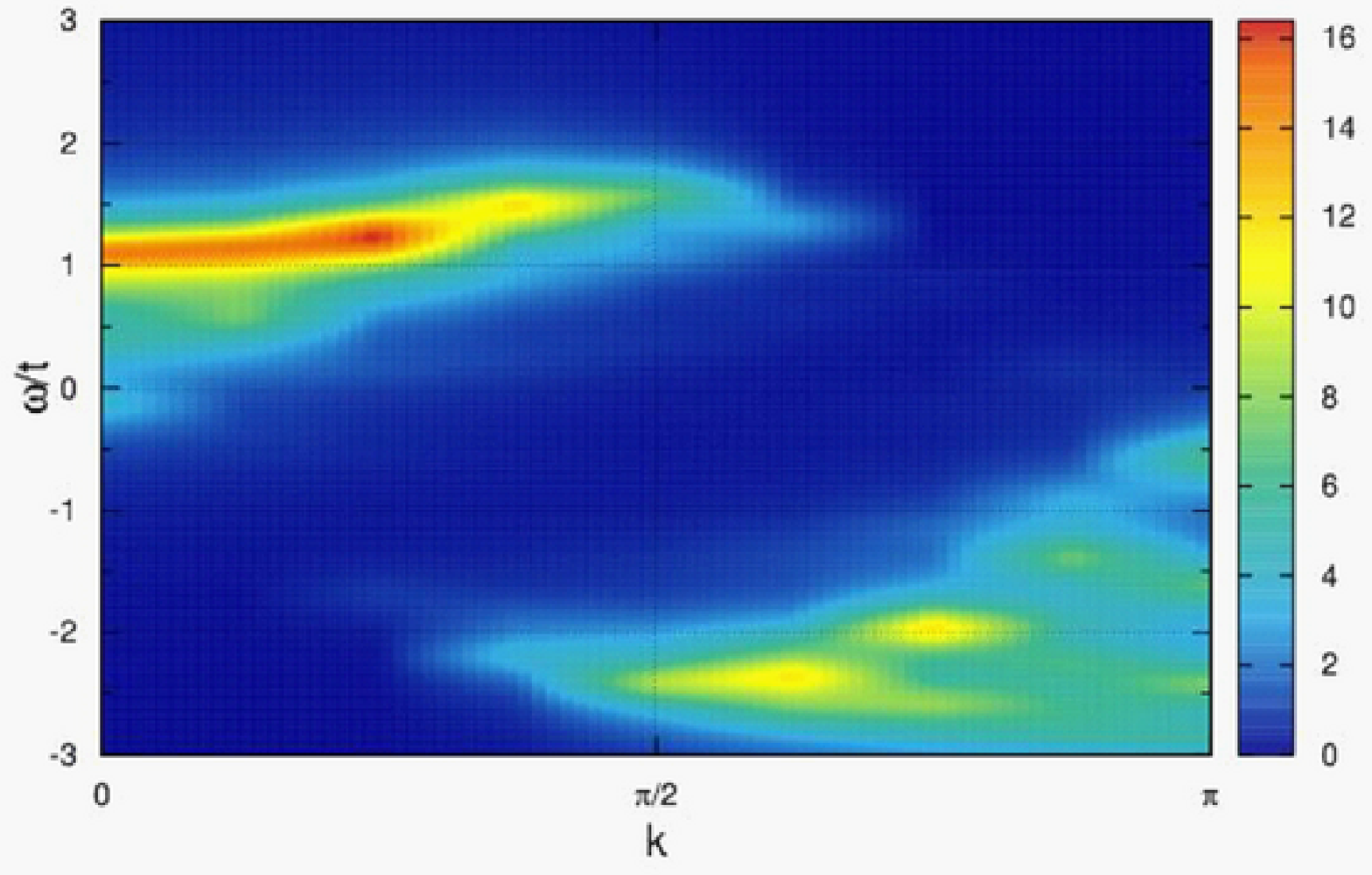}
\end{center}
\caption{Angle-resolved spectrum for $\Psi_{2}=\vec{n}\cdot\vec{\sigma}\xi^{\alpha}$}
\label{fig1b}
\end{figure}

\begin{figure}[htbp]
\begin{center}
\includegraphics[width=8.5cm]{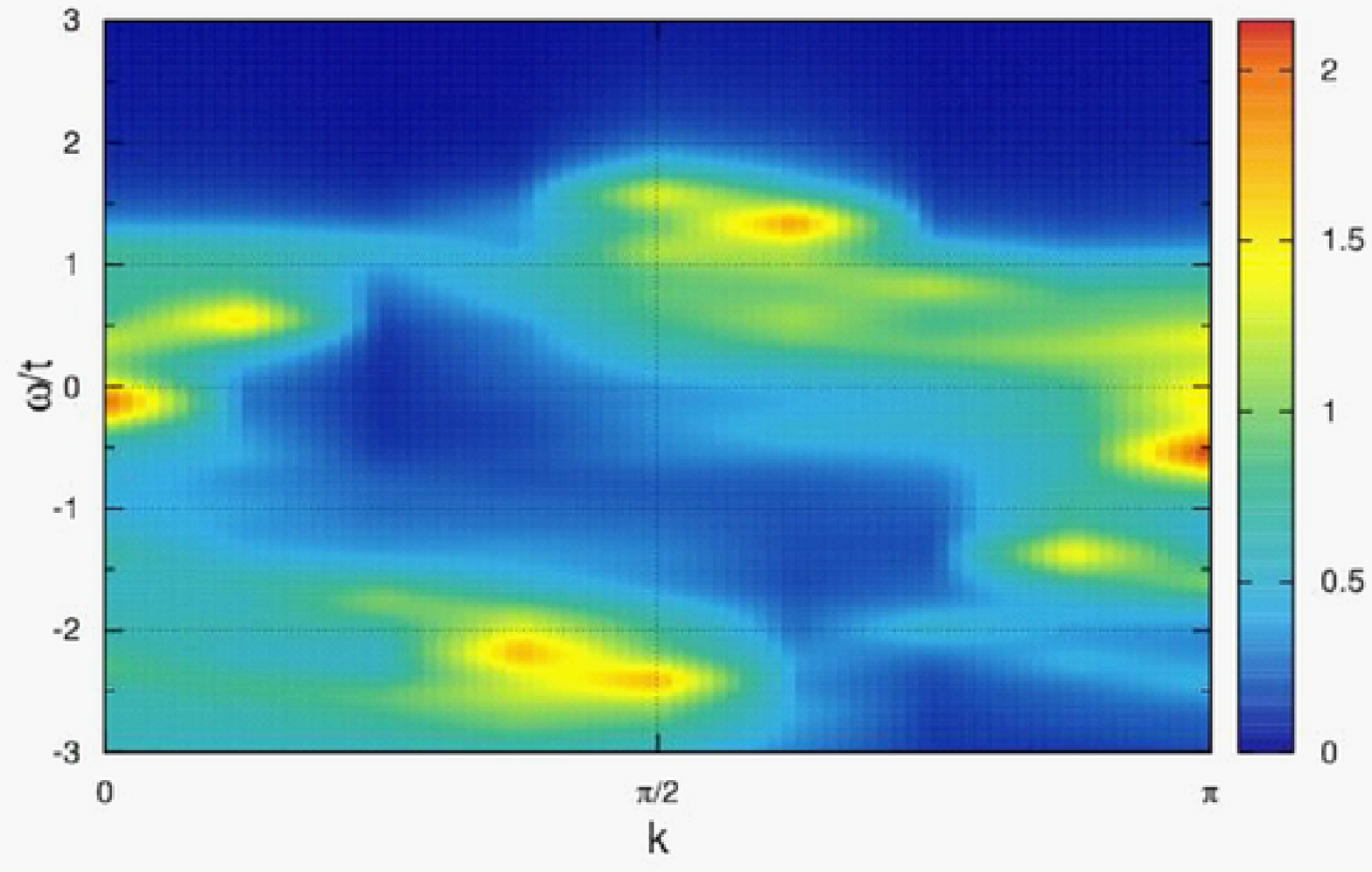}
\end{center}
\caption{Angle-resolved spectrum for $\Psi_{3}^{-}=\vec{n}\cdot\vec{\sigma}\xi^{\alpha\alpha}$}
\label{fig1c}
\end{figure}

\begin{figure}[htbp]
\begin{center}
\includegraphics[width=8.5cm]{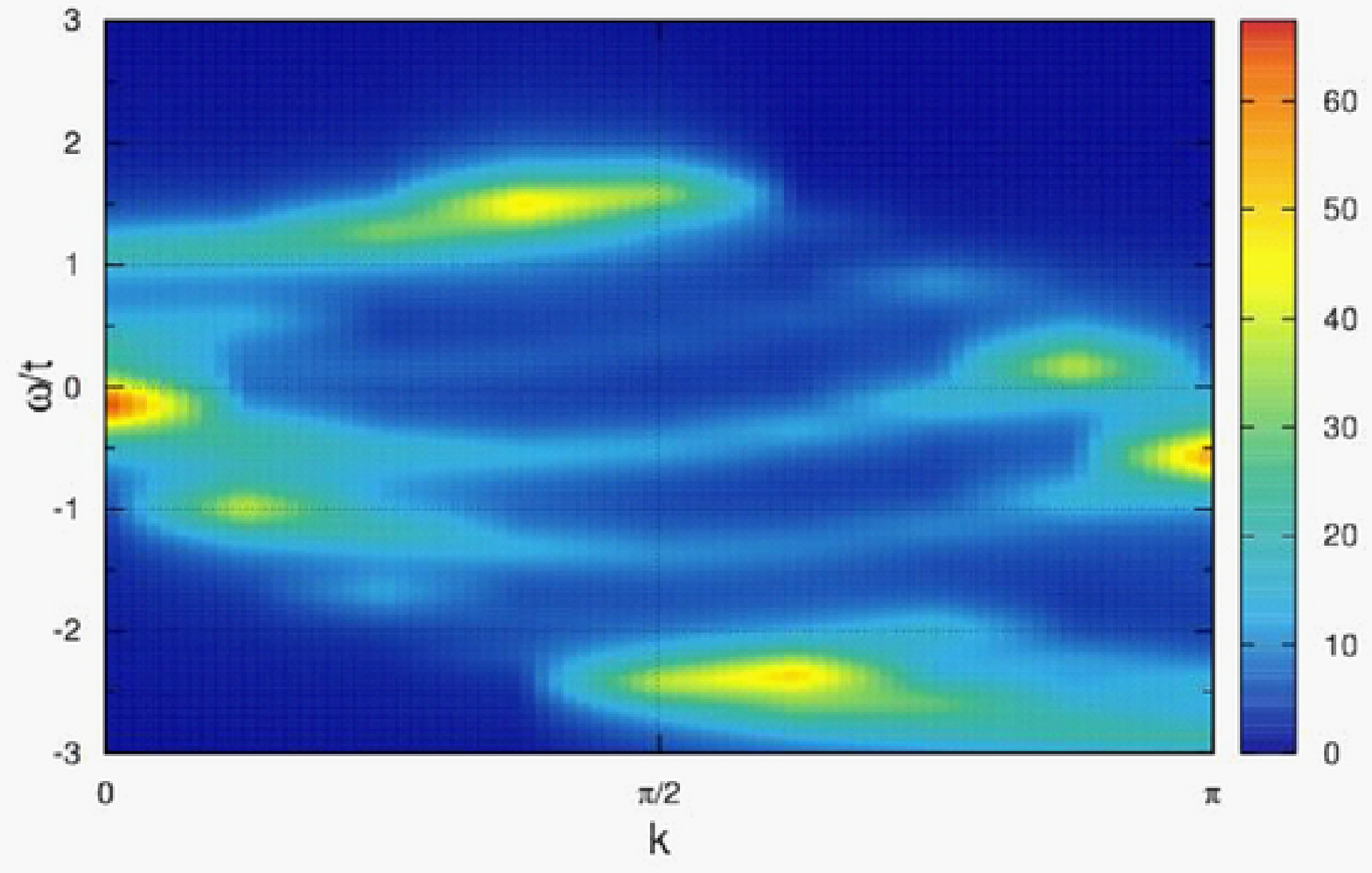}
\end{center}
\caption{Angle-resolved spectrum for $\Psi_{3}^{+}=\vec{n}\cdot\vec{\sigma}\left(\vec{n}\cdot\vec{\sigma}\xi^{\alpha}\right)^{\alpha}$}
\label{fig1d}
\end{figure}

\begin{figure}[htbp]
\begin{center}
\includegraphics[width=8.5cm]{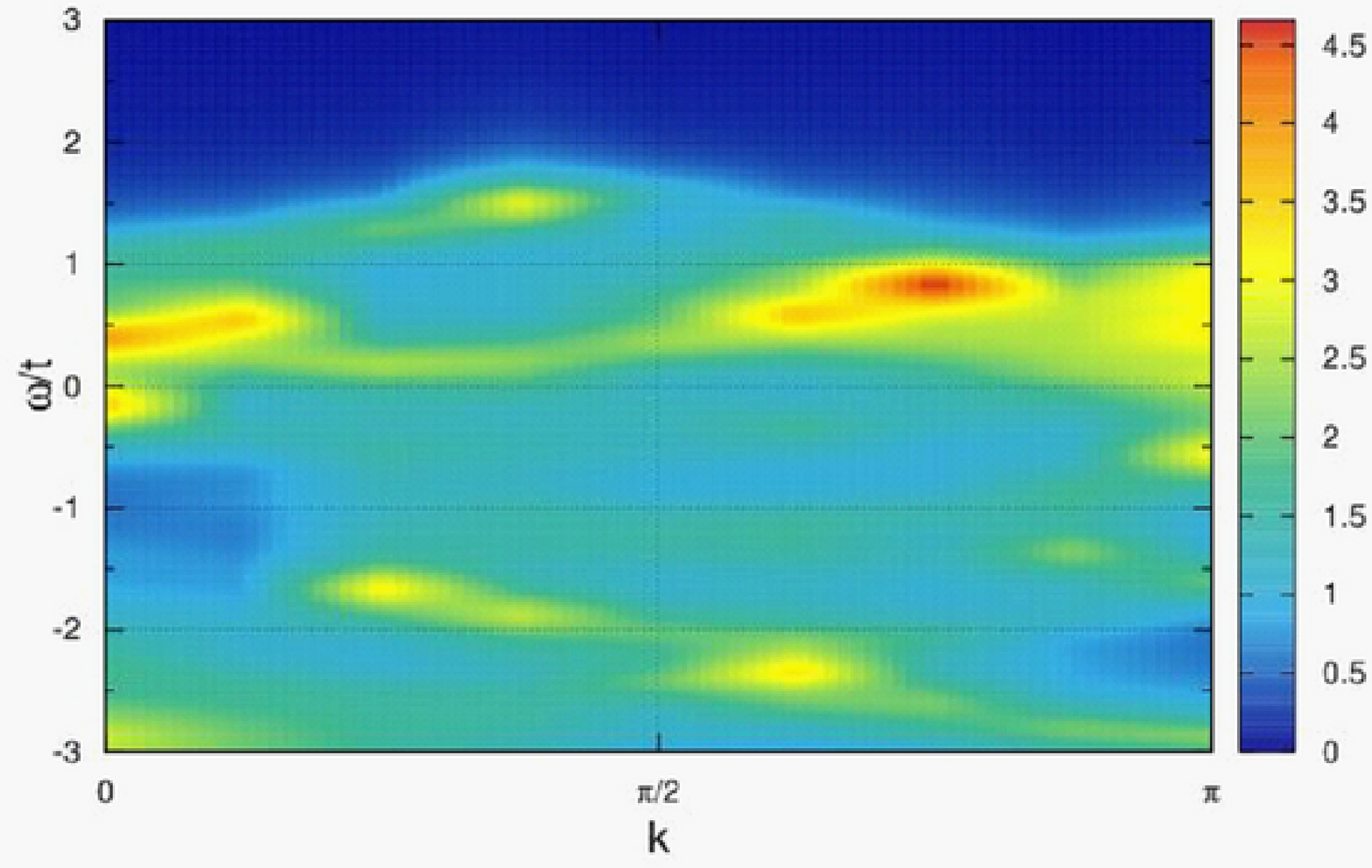}
\end{center}
\caption{Angle-resolved spectrum for $\Psi_{4}^{--}$}
\label{fig1e}
\end{figure}

\begin{figure}[htbp]
\begin{center}
\includegraphics[width=8.5cm]{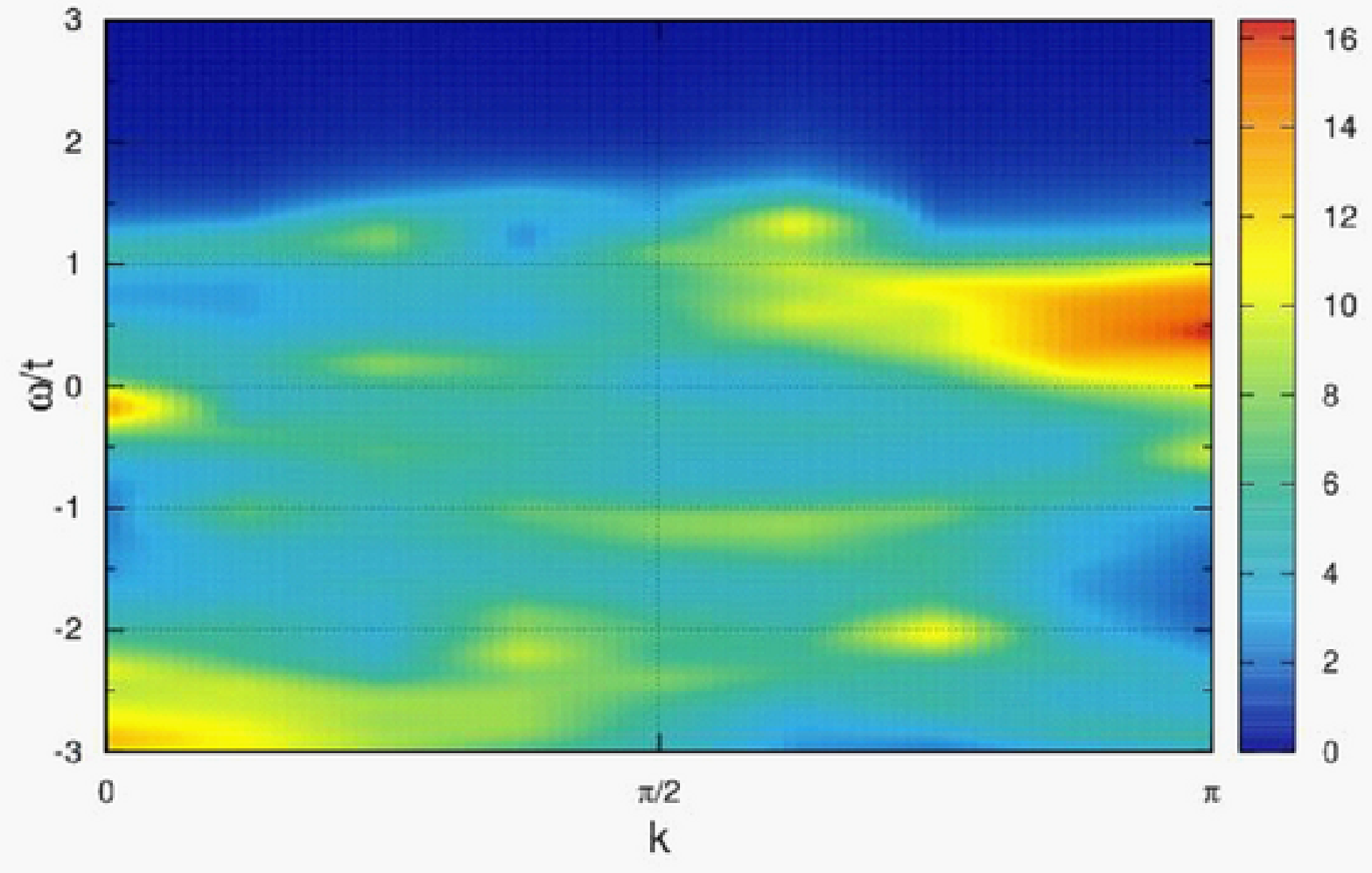}
\end{center}
\caption{Angle-resolved spectrum for $\Psi_{4}^{-+}$}
\label{fig1f}
\end{figure}

\begin{figure}[htbp]
\begin{center}
\includegraphics[width=8.5cm]{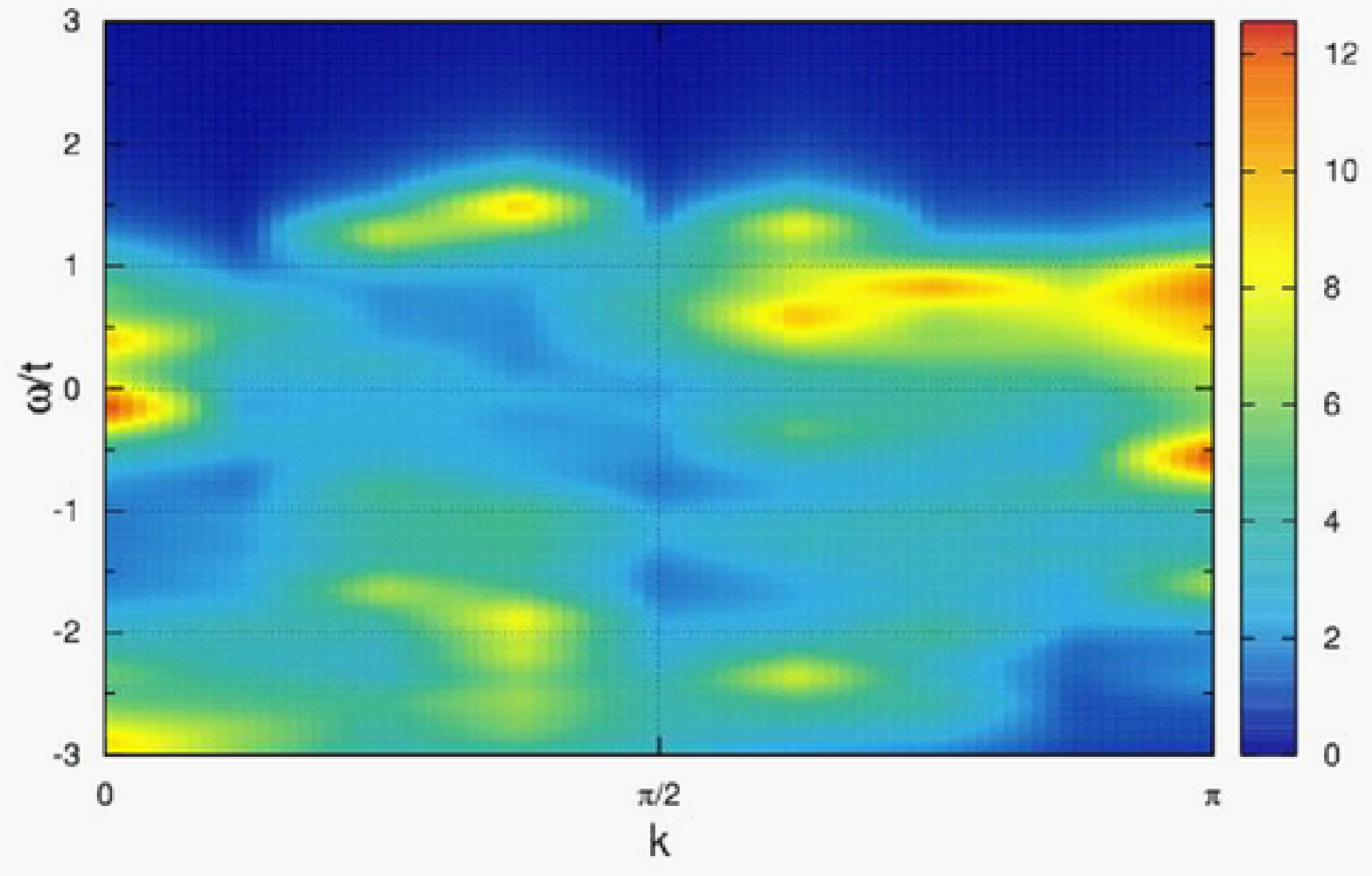}
\end{center}
\caption{Angle-resolved spectrum for $\Psi_{4}^{+-}$}
\label{fig1g}
\end{figure}

\begin{figure}[htbp]
\begin{center}
\includegraphics[width=8.5cm]{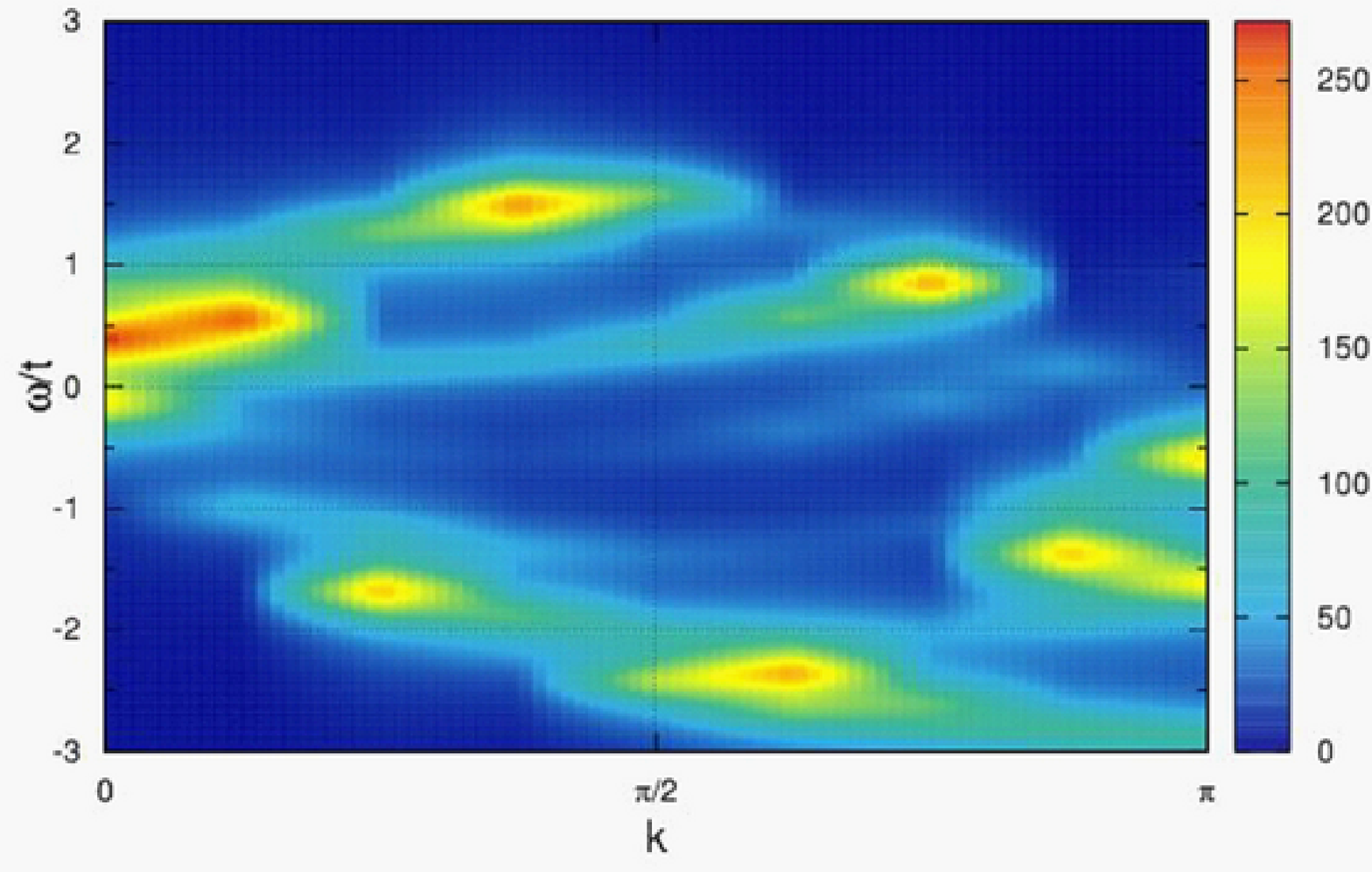}
\end{center}
\caption{Angle-resolved spectrum for $\Psi_{4}^{++}$}
\label{fig1h}
\end{figure}

\begin{figure}[htbp]
\begin{center}
\includegraphics[width=8cm]{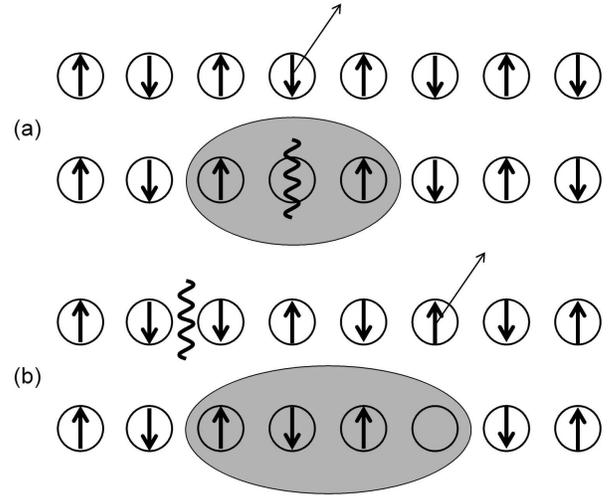}
\end{center}
\caption{Photoemission processes from AF-like background: (a) the electron removal process by $\Psi_{2}$ and (b) the electron removal process by $\Psi_{4}^{++}$.}
\label{fig1i}
\end{figure}

Let us first examine the half-filled case. We present the angle-resolved spectrum for each component of $\Psi$ in Figs.~\ref{fig1a}-\ref{fig1h}. The figure~\ref{fig1a} for $\Psi_{1}=\xi$ corresponds to the standard ARPES result. Our numerical result of the ARPES spectrum is consistent with the past analytical and numerical results, and we clearly observe the spinon and holon branches with different energy scales $J$ and $t$, respectively. Based on this consistency, let us particularly focus on the spectra for $\Psi_{2}$, $\Psi_{3}^{+}$, and $\Psi_{4}^{++}$ in the following sentences. The doped holes in these modes strongly couple with various sizes of the spin strings. Thus, they may not be simple single-particle-type excitations, although they are still fermionic excitations. We expect that they would play crucial roles on the formation of collective excitations such as spinon and holon excitations.

At first, we examine the spectrum for $\Psi_{2}=\vec{n}\cdot\vec{\sigma}\xi^{\alpha}$ in Fig.~\ref{fig1b}. We find that the main band exactly traces the spinon branch. This is quite unique, since such a property enables us to do more advanced spectroscopy in the sense that we can select one structure of complicated bands in correlated electron systems. The reason for such separable spectroscopy is quite simple. When we remove one electron from the AF insulating ground state, the domain boundary (spinon) of the AF state is created at around the position where the electron is removed. Since the electron removal by the operator $\Psi_{2}$ counts the change in the near-neighbor spin structure as well as the created hole, the operator necessarily presents the spinon creation [see Fig.~\ref{fig1i}(a)]. Therefore, we can regard $\Psi_{2}$ as '{\it the spinon operator}'. Note that this is different from the spinon operator $f$ in the context of the slave boson representation $\xi=fb^{\dagger}$.

On the other hand in Figs.~\ref{fig1d} and \ref{fig1h}, the main bands for $\Psi_{3}^{+}$ and $\Psi_{4}^{++}$ trace the holon branch, and the spectral weight of the spinon branch is not strong. Thus, they are basically '{\it the holon operators}'. It is noted that the covering of the holon branch by these types of operators becomes much better as the spin string becomes longer. The readers may wonder why they behave like the holon excitation, although they are dressed with many spins. The answer may originate in a fact that the spin string attached to $\Psi_{3}^{+}$ and $\Psi_{4}^{++}$ can regulate the spin structure after the photoemission so that the resulting final state has just one hole. Let us consider the situation in which they are operated to the AF-like state, but the state contains one spinon bases owing to quantum fluctuation. Then, the operation of $\Psi_{3}^{+}$ and $\Psi_{4}^{++}$ to the state can remove the spinon as well as creation of the hole [see Fig.~\ref{fig1i}(b)].

These results suggest that each branch in the ARPES data originates from a particular composite excitation. Interestingly, these spectral intensities are much stronger than that of the single-particle spectrum, and there is only weak continuum part. This weak continuum feature really represents that these composite excitations are more important than the simple $\xi$ state.

In comparison with $\Psi_{2}$, $\Psi_{3}^{+}$, and $\Psi_{4}^{++}$, the spectra for $\Psi_{3}^{-}$, $\Psi_{4}^{--}$, $\Psi_{4}^{-+}$, and $\Psi_{4}^{+-}$ as shown in Figs.~\ref{fig1c}, \ref{fig1e}, \ref{fig1f}, and \ref{fig1g} are more scattered, though a considerable amount of spectral intensity is still on the holon branch at $\pi/2\le k\le\pi$. For these excitations, the multiple spin excitations are away from a created hole by the photoemission, thus they seem to be somehow different from $\Psi_{2}$, $\Psi_{3}^{+}$, and $\Psi_{4}^{++}$.

\subsection{Hole doping}

\begin{figure}[htbp]
\begin{center}
\includegraphics[width=8.5cm]{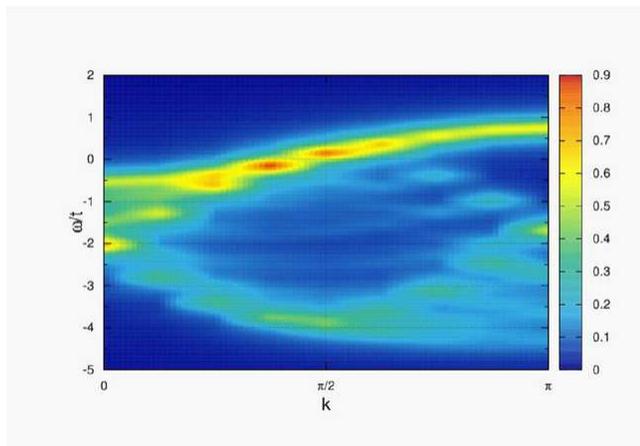}
\end{center}
\caption{ARPES spectrum (spectrum for the bare fermionic particle $\Psi_{1}$)}
\label{fig2a}
\end{figure}

\begin{figure}[htbp]
\begin{center}
\includegraphics[width=8.5cm]{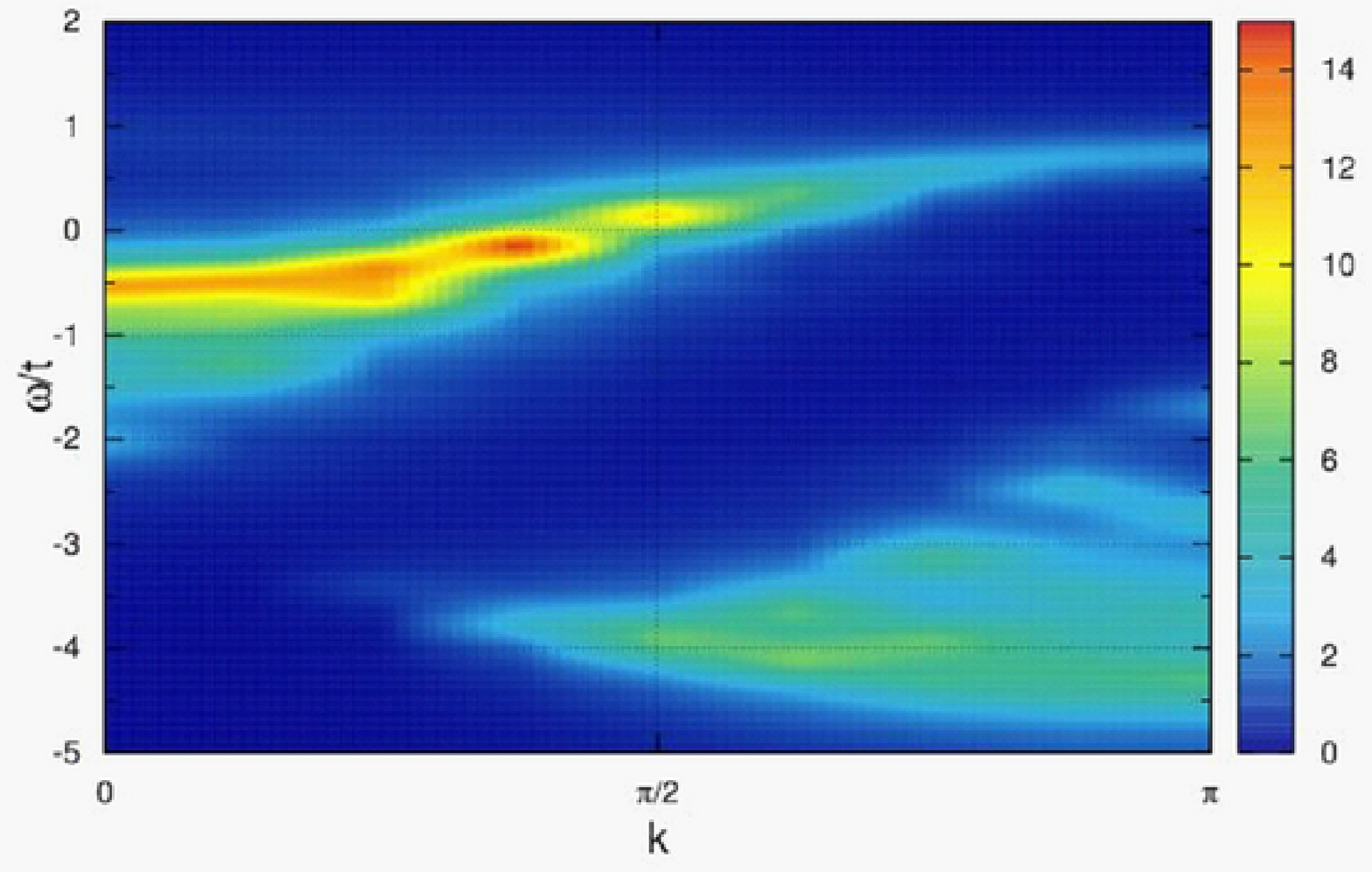}
\end{center}
\caption{Angle-resolved spectrum for $\Psi_{2}=\vec{n}\cdot\vec{\sigma}\xi^{\alpha}$}
\label{fig2b}
\end{figure}

\begin{figure}[htbp]
\begin{center}
\includegraphics[width=8.5cm]{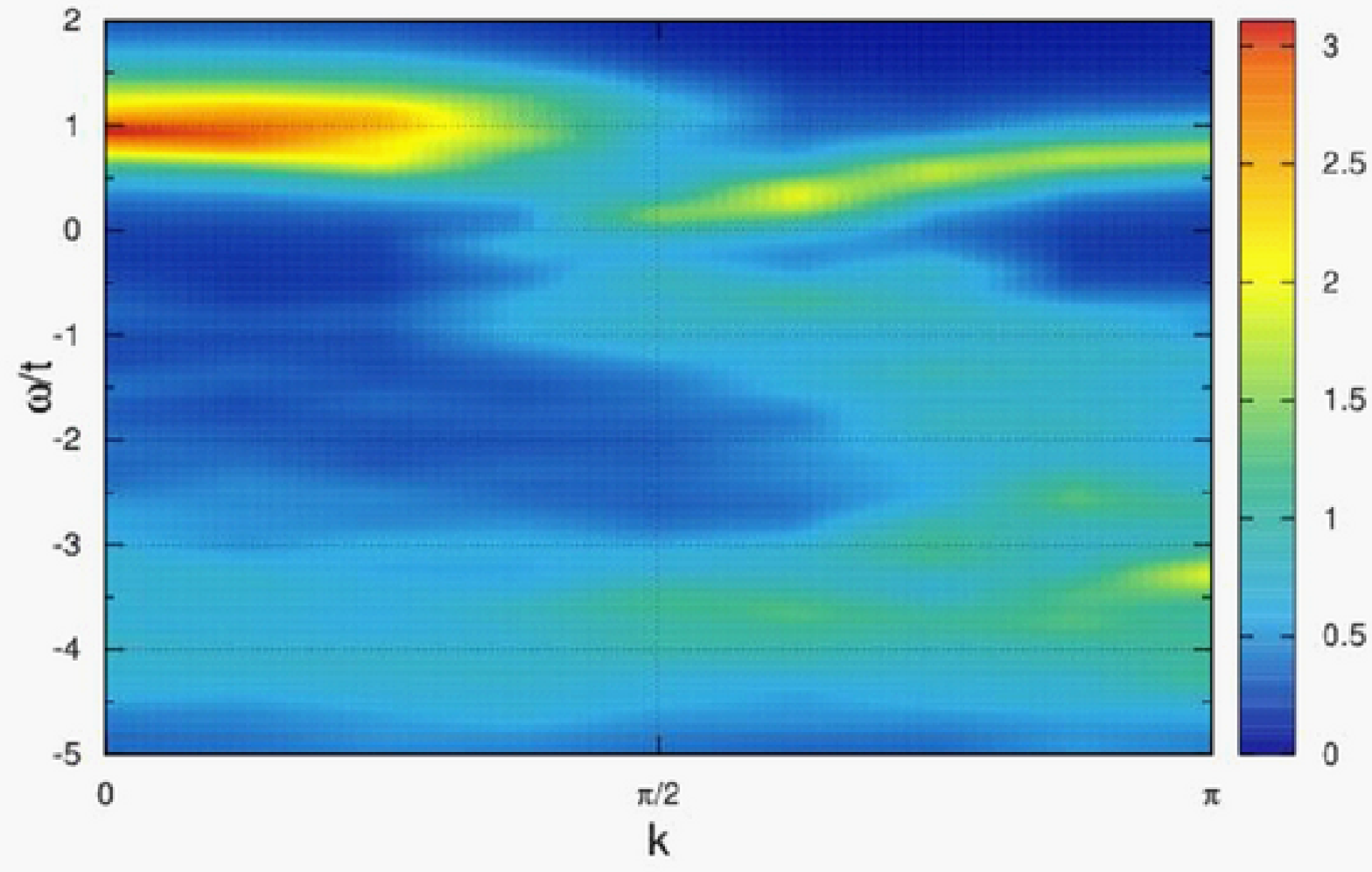}
\end{center}
\caption{Angle-resolved spectrum for $\Psi_{3}^{-}=\vec{n}\cdot\vec{\sigma}\xi^{\alpha\alpha}$}
\label{fig2c}
\end{figure}

\begin{figure}[htbp]
\begin{center}
\includegraphics[width=8.5cm]{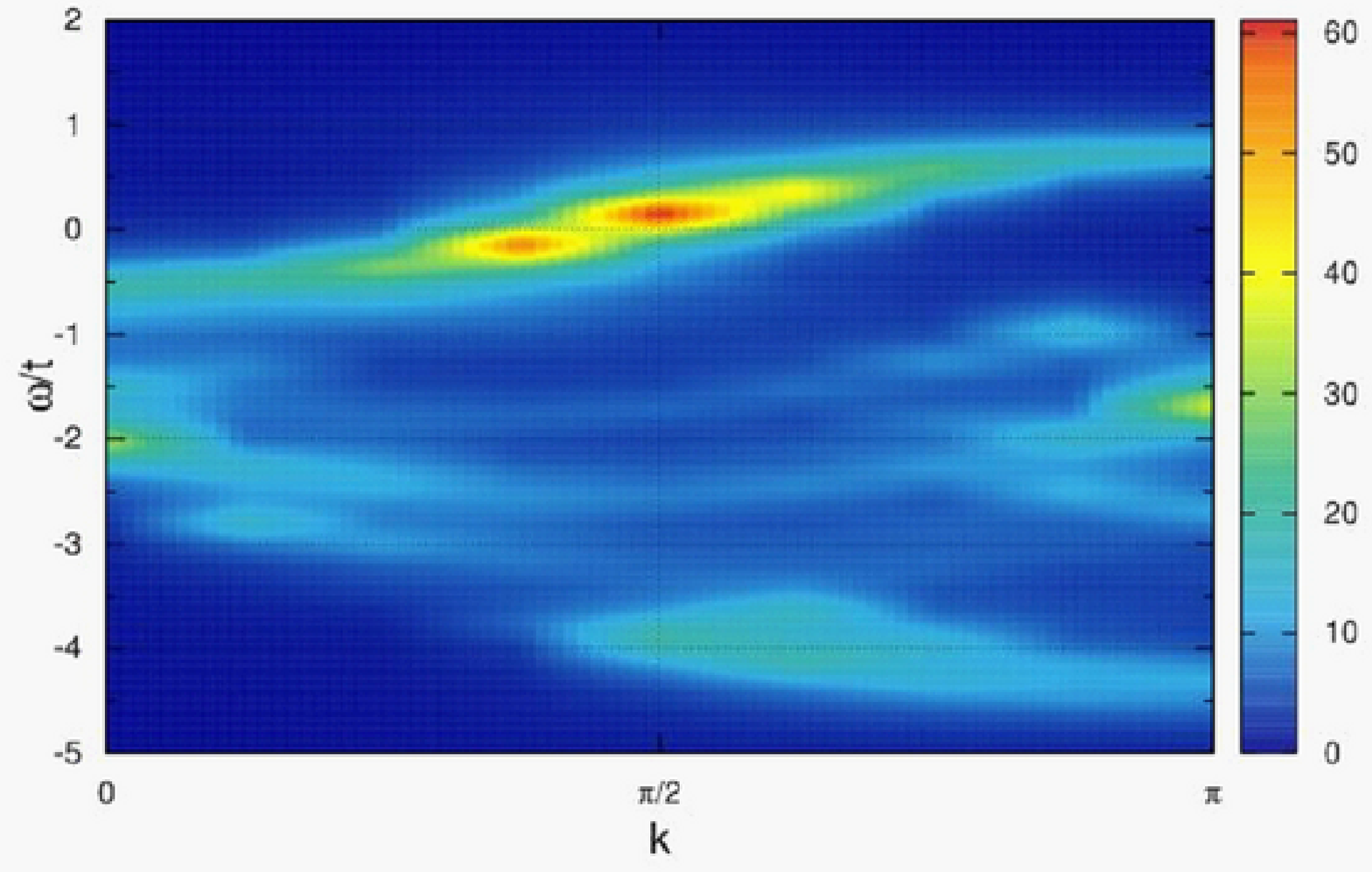}
\end{center}
\caption{Angle-resolved spectrum for $\Psi_{3}^{+}=\vec{n}\cdot\vec{\sigma}\left(\vec{n}\cdot\vec{\sigma}\xi^{\alpha}\right)^{\alpha}$}
\label{fig2d}
\end{figure}

\begin{figure}[htbp]
\begin{center}
\includegraphics[width=8.5cm]{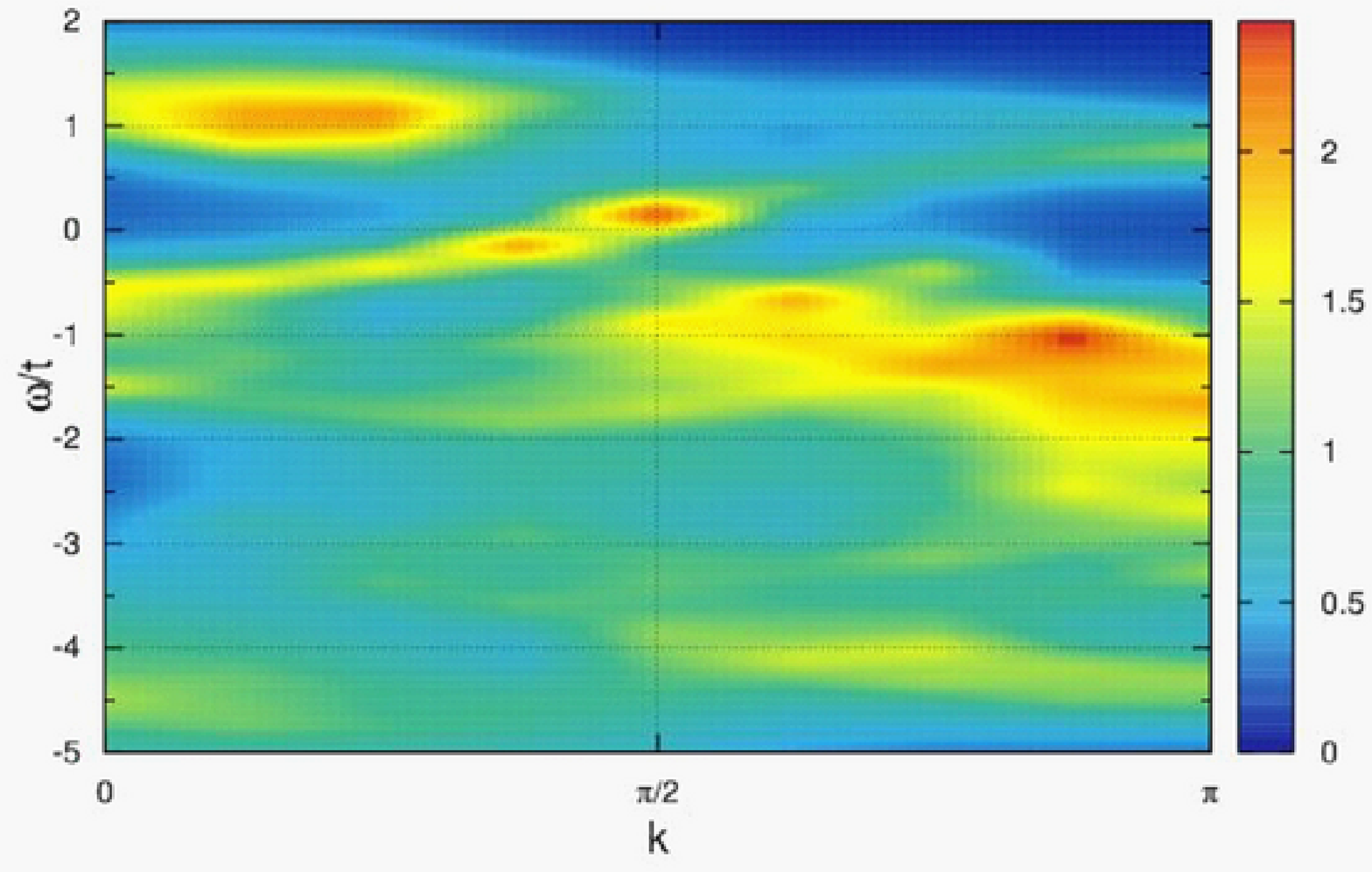}
\end{center}
\caption{Angle-resolved spectrum for $\Psi_{4}^{--}$}
\label{fig2e}
\end{figure}

\begin{figure}[htbp]
\begin{center}
\includegraphics[width=8.5cm]{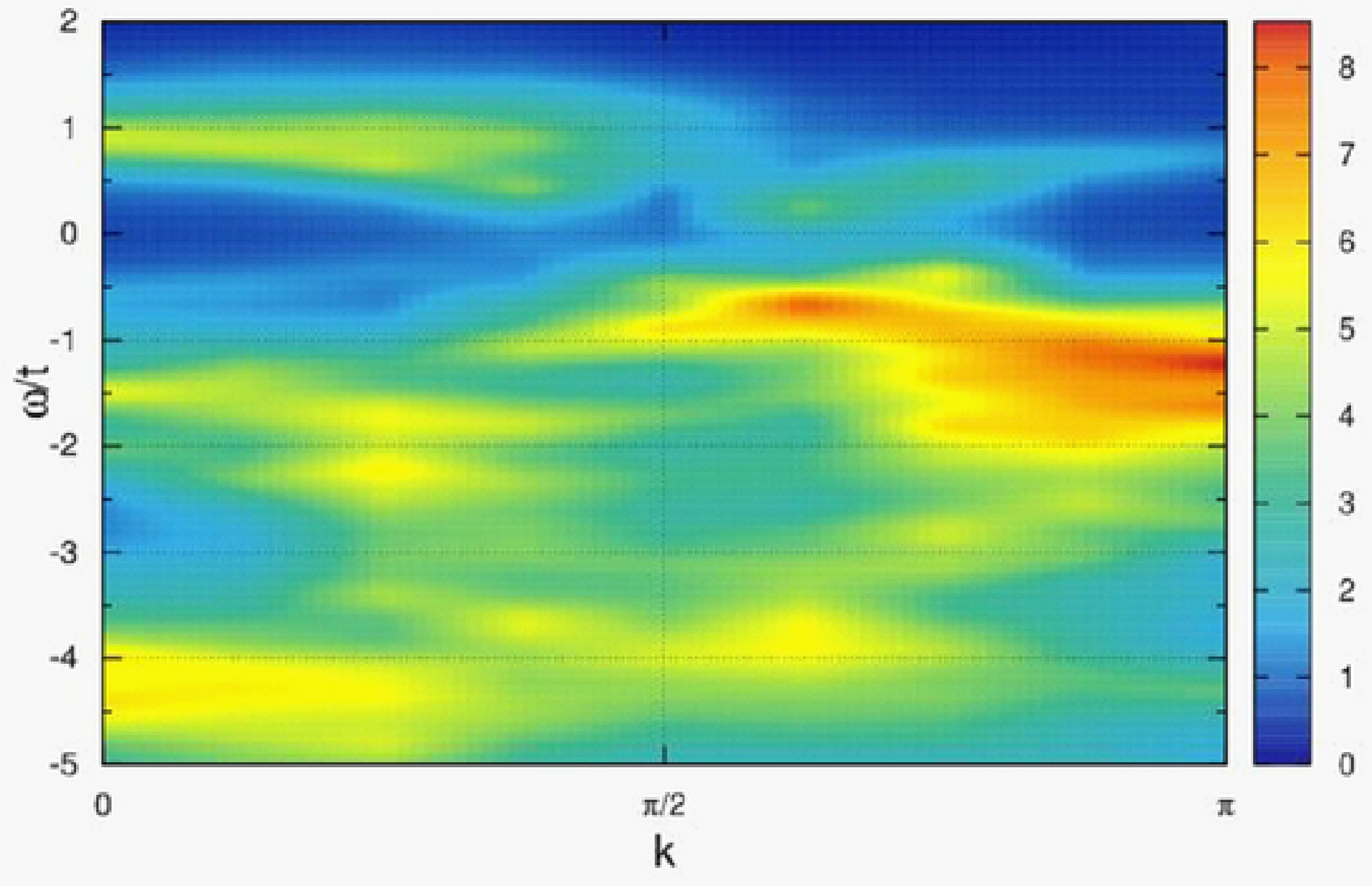}
\end{center}
\caption{Angle-resolved spectrum for $\Psi_{4}^{-+}$}
\label{fig2f}
\end{figure}

\begin{figure}[htbp]
\begin{center}
\includegraphics[width=8.5cm]{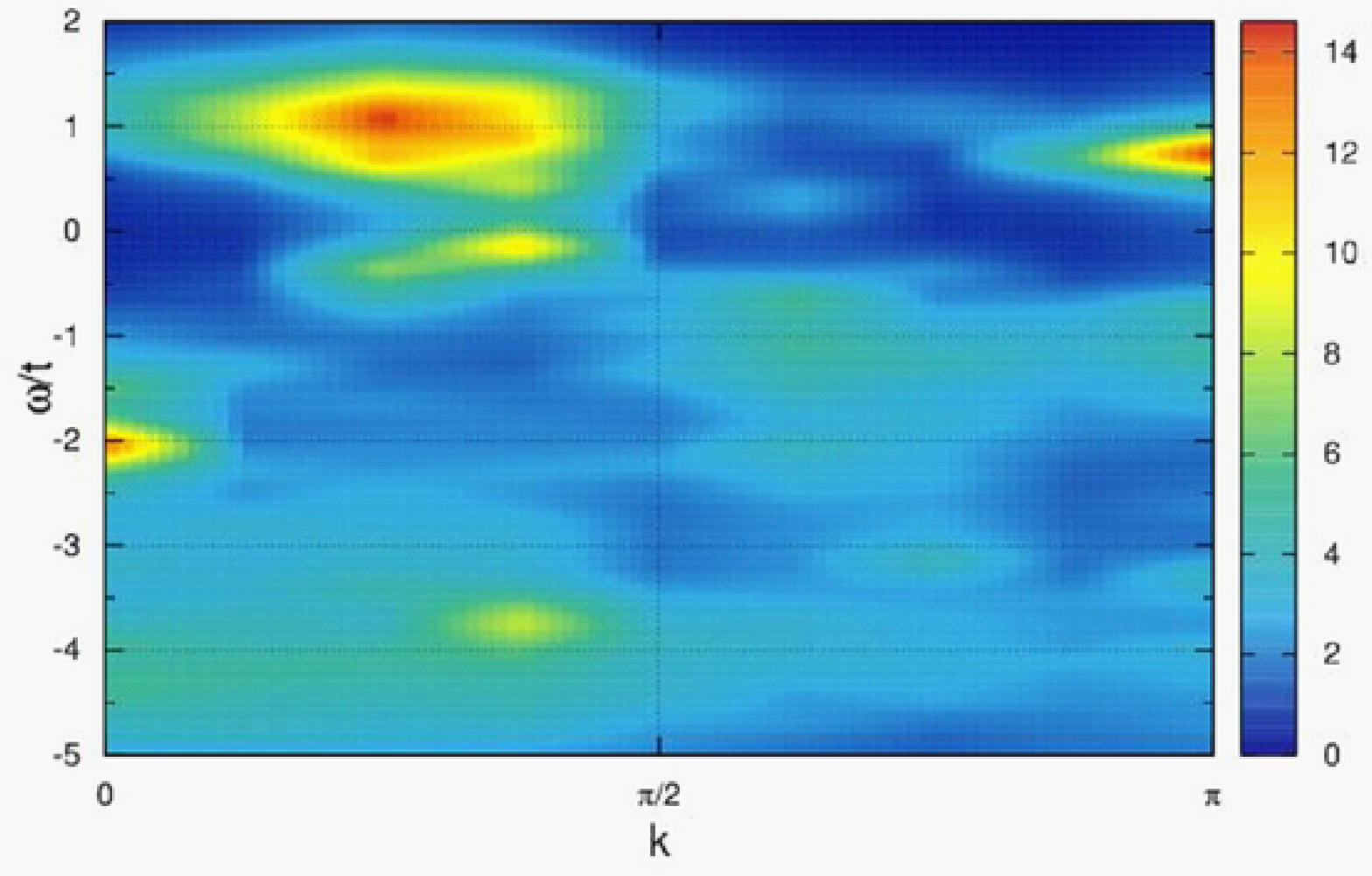}
\end{center}
\caption{Angle-resolved spectrum for $\Psi_{4}^{+-}$}
\label{fig2g}
\end{figure}

\begin{figure}[htbp]
\begin{center}
\includegraphics[width=8.5cm]{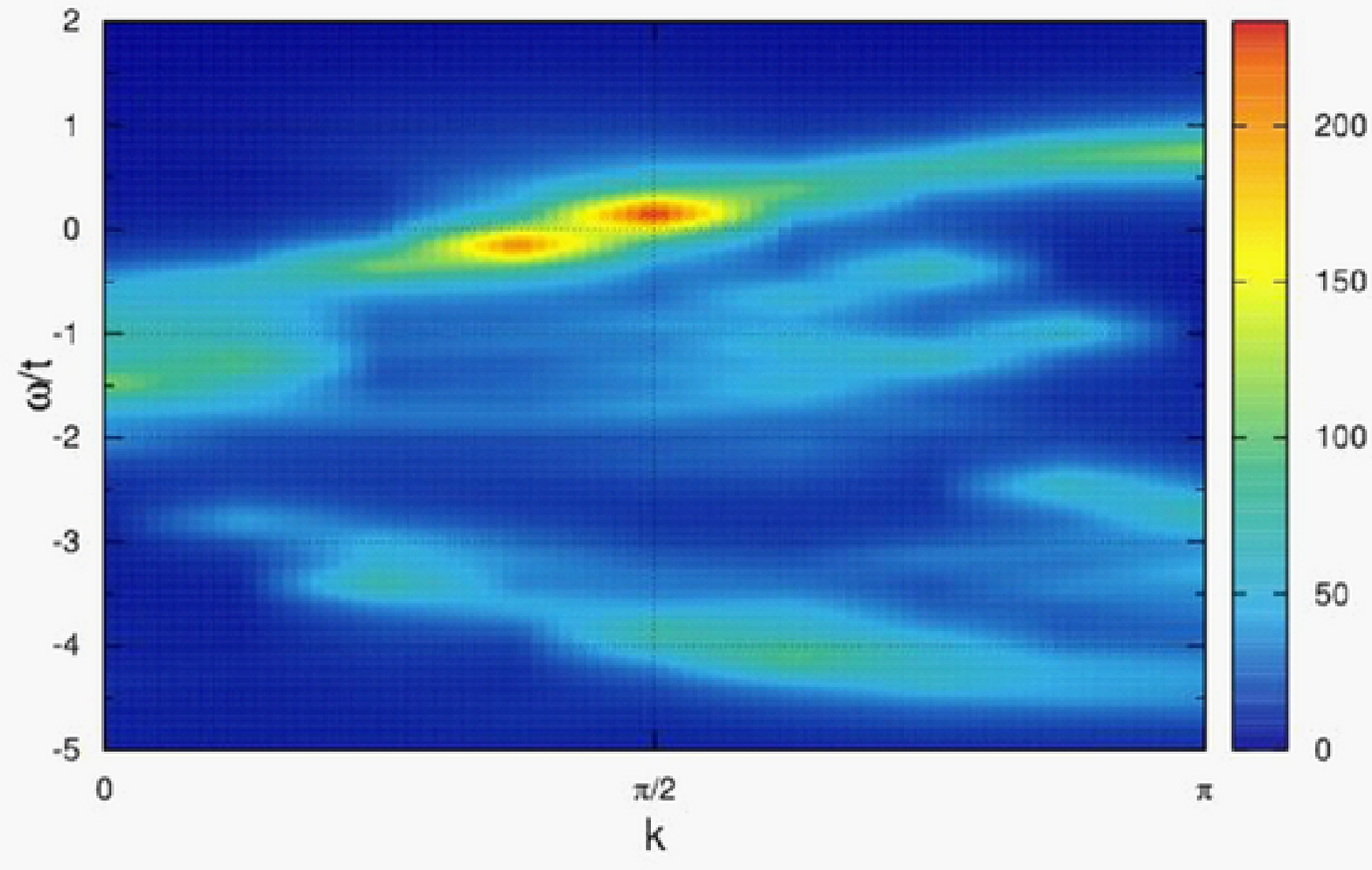}
\end{center}
\caption{Angle-resolved spectrum for $\Psi_{4}^{++}$}
\label{fig2h}
\end{figure}

Next we examine the doped case in Figs.~\ref{fig2a}-\ref{fig2h}. In this case also, the important composite excitations near the Fermi level are $\Psi_{2}$, $\Psi_{3}^{+}$, and $\Psi_{4}^{++}$. We find that the spectral intensity concentrates on $k=\pi/2$ and $\omega/t=0$ as the doped hole is dressed with longer spin string. At this momentum, the spinon and holon branches merge together. Thus, the angle-resolved spectra for $\Psi_{3}^{+}$ and $\Psi_{4}^{++}$ have enough information for real electronic excitation as well as the phase string that connects spinon with holon.

In the standard ARPES spectrum (Fig.~\ref{fig2a}), we observe the metallic branch showing the Tomonaga-Luttinger liquid behavior. The spectral weight above the Fermi level originates in the holon branch at $k>\pi/2$. However, for the composite excitations $\Psi_{2}$, $\Psi_{3}^{+}$, and $\Psi_{4}^{++}$, the spectral intensity of this branch tends to be vanishing above the Fermi level. In particular, the spectral intensity for $\Psi_{3}^{+}$ and $\Psi_{4}^{++}$ is strong only at the Fermi level. On the other hand, the spectrum for $\Psi_{3}^{-}$ has considerable weight above the Fermi level. Therefore, the composite excitations behave quite differently from the fundamental excitation $\Psi_{1}=\xi$. Although we observe single metallic band in the ARPES spectrum, these results suggest that the character of the wave function changes at the Fermi level. This is because for instance $\Psi_{2}$ corresponds to the self-energy correction to $\Psi_{1}$ in the standard field-theoretical terminology, and this means that the damping mechanism of the fundamental excitation strongly depends on the momentum. We will later discuss about this feature again in comparison with the presence of the pseudogap in 2D cases and high-$T_{c}$ cuprates.

The spectra for $\Psi_{3}^{-}$, $\Psi_{4}^{--}$, $\Psi_{4}^{-+}$, and $\Psi_{4}^{+-}$ seem to contain different information. Actually, we observe relatively strong spectral intensity at $0\le k\le\pi$ and $\omega/t\sim 1$. This shadow band seems to naturally connect to the holon branch at $\pi/2\le k\le\pi$, and intersects with the metallic band at around $k=\pi/2$. Owing to this intersection, the band seems to split into two substructures. Later, we will also discuss more about this intersection in connection with the presence of the pseudogap in high-$T_{c}$ cuprates.

\section{Numerical Results II: 2D case}

\begin{figure}[htbp]
\begin{center}
\includegraphics[width=8.5cm]{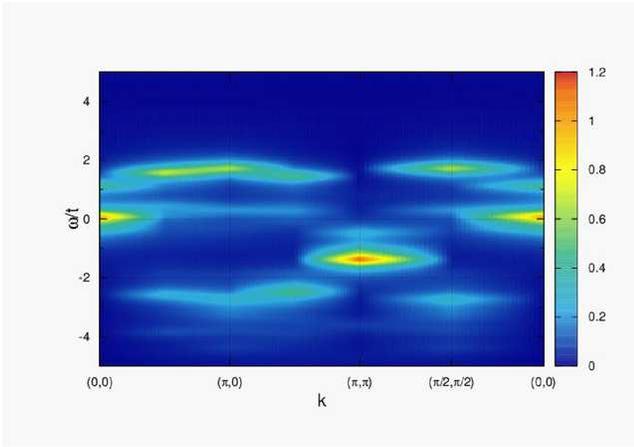}
\end{center}
\caption{ARPES spectrum (spectrum for $\Psi_{1}=\xi$) at half-filling}
\label{fig3a}
\end{figure}

\begin{figure}[htbp]
\begin{center}
\includegraphics[width=8.5cm]{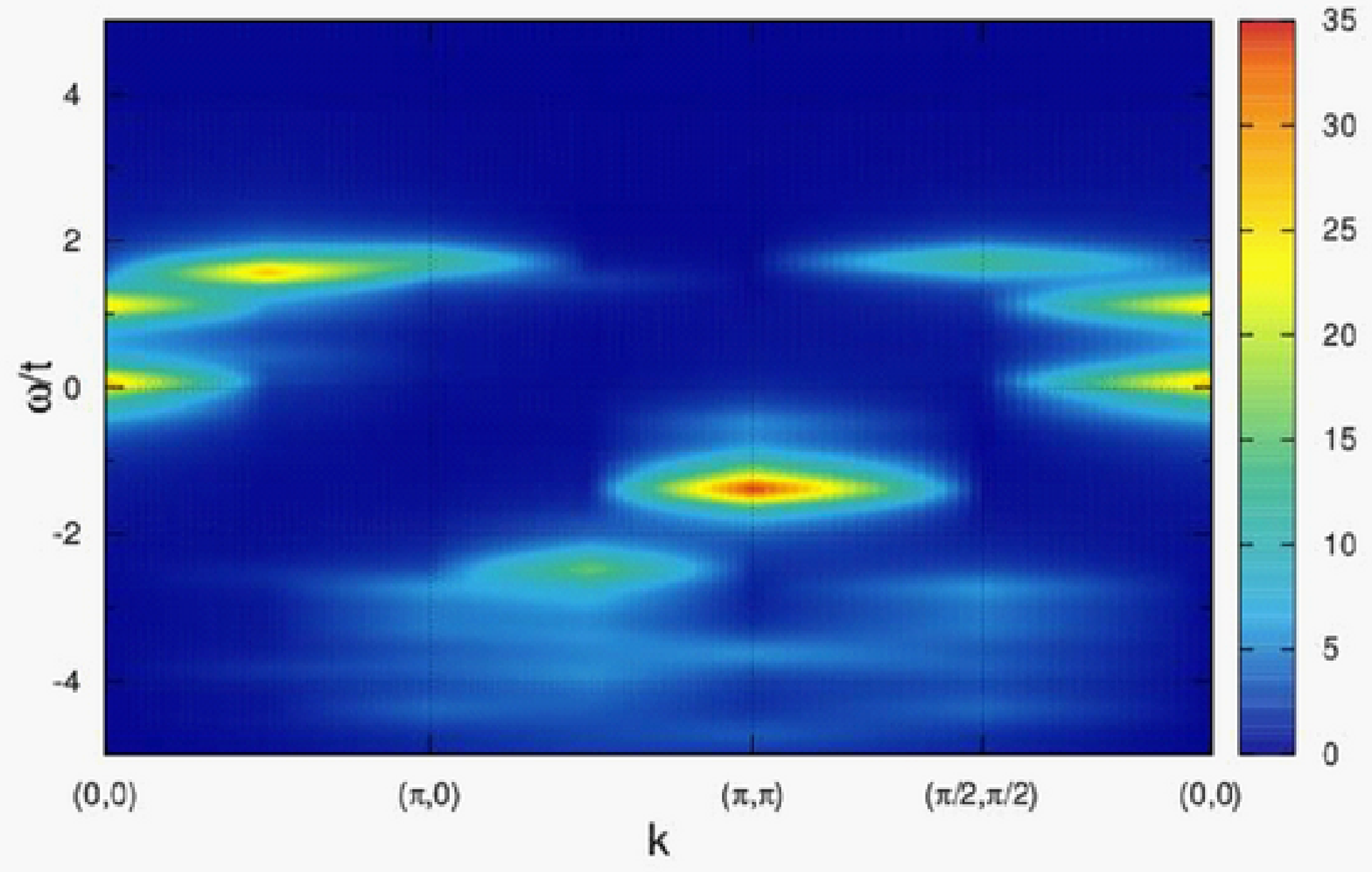}
\end{center}
\caption{Spectrum for $\Psi_{2}=\vec{n}\cdot\vec{\sigma}\xi^{\alpha}$ at half-filling}
\label{fig3b}
\end{figure}

\begin{figure}[htbp]
\begin{center}
\includegraphics[width=8.5cm]{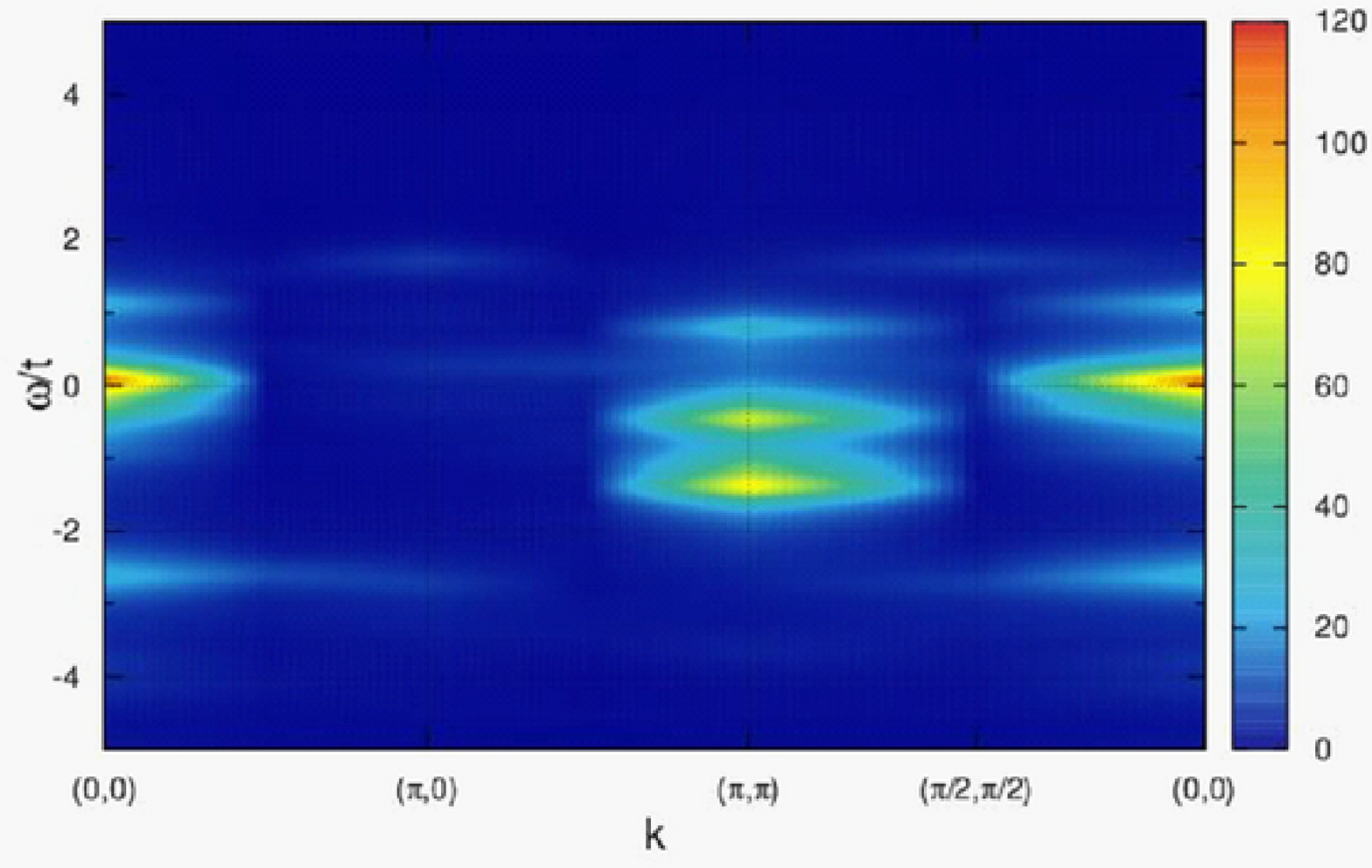}
\end{center}
\caption{Spectrum for $\Psi_{3}^{-}=\vec{n}\cdot\vec{\sigma}\xi^{\alpha\alpha}$ at half-filling}
\label{fig3c}
\end{figure}

\begin{figure}[htbp]
\begin{center}
\includegraphics[width=8.5cm]{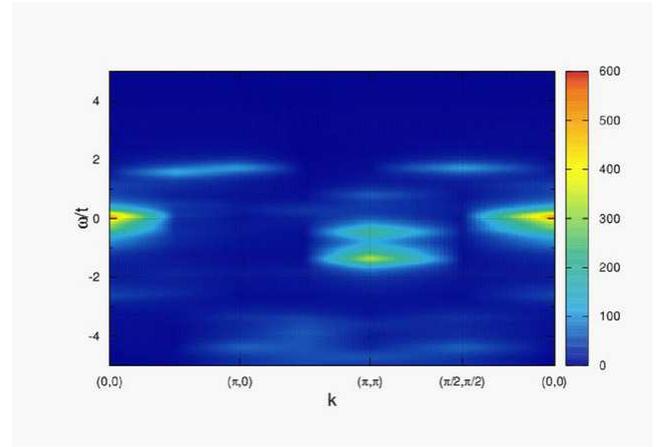}
\end{center}
\caption{Spectrum for $\Psi_{3}^{+}=\vec{n}\cdot\vec{\sigma}\left(\vec{n}\cdot\vec{\sigma}\xi^{\alpha}\right)^{\alpha}$ at half-filling}
\label{fig3d}
\end{figure}

\begin{figure}[htbp]
\begin{center}
\includegraphics[width=8.5cm]{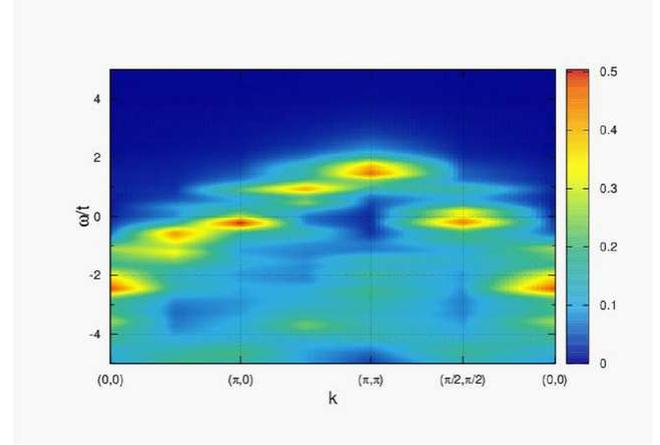}
\end{center}
\caption{ARPES spectrum (spectrum for $\Psi_{1}=\xi$) for the two-hole state}
\label{fig3e}
\end{figure}

\begin{figure}[htbp]
\begin{center}
\includegraphics[width=8.5cm]{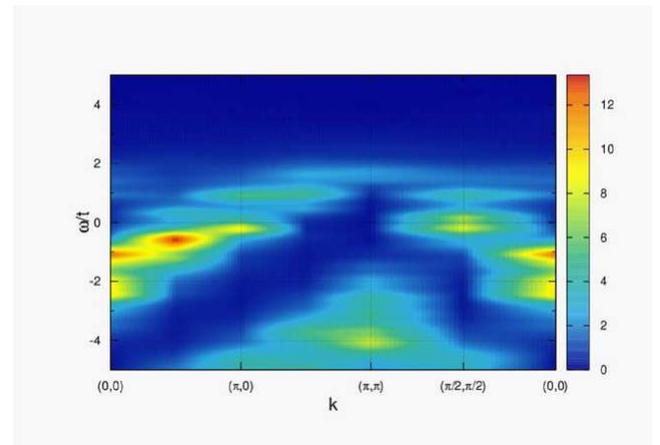}
\end{center}
\caption{Spectrum for $\Psi_{2}=\vec{n}\cdot\vec{\sigma}\xi^{\alpha}$ for the two-hole state}
\label{fig3f}
\end{figure}

\begin{figure}[htbp]
\begin{center}
\includegraphics[width=8.5cm]{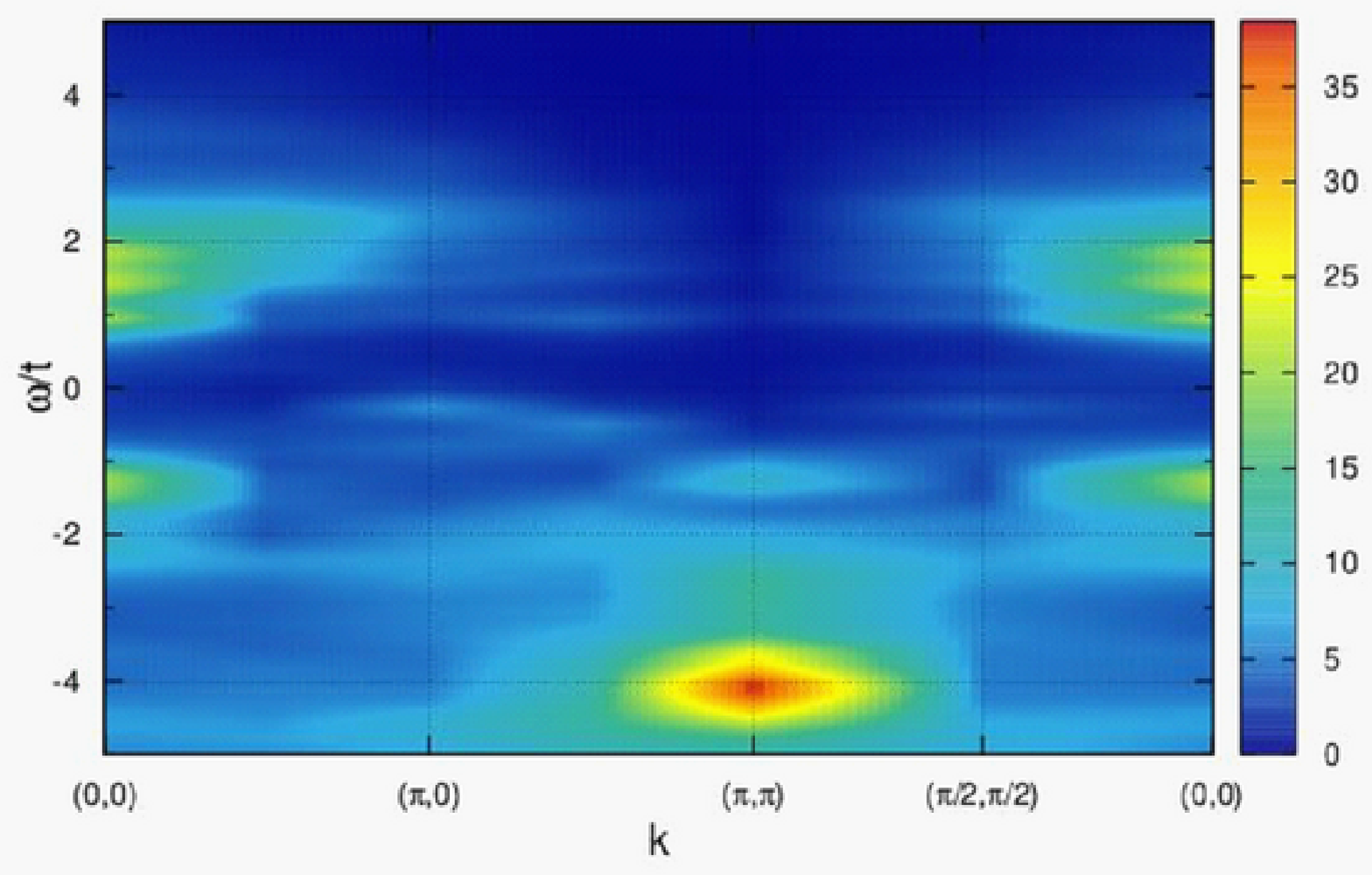}
\end{center}
\caption{Spectrum for $\Psi_{3}^{-}=\vec{n}\cdot\vec{\sigma}\xi^{\alpha\alpha}$ for the two-hole state}
\label{fig3g}
\end{figure}

\begin{figure}[htbp]
\begin{center}
\includegraphics[width=8.5cm]{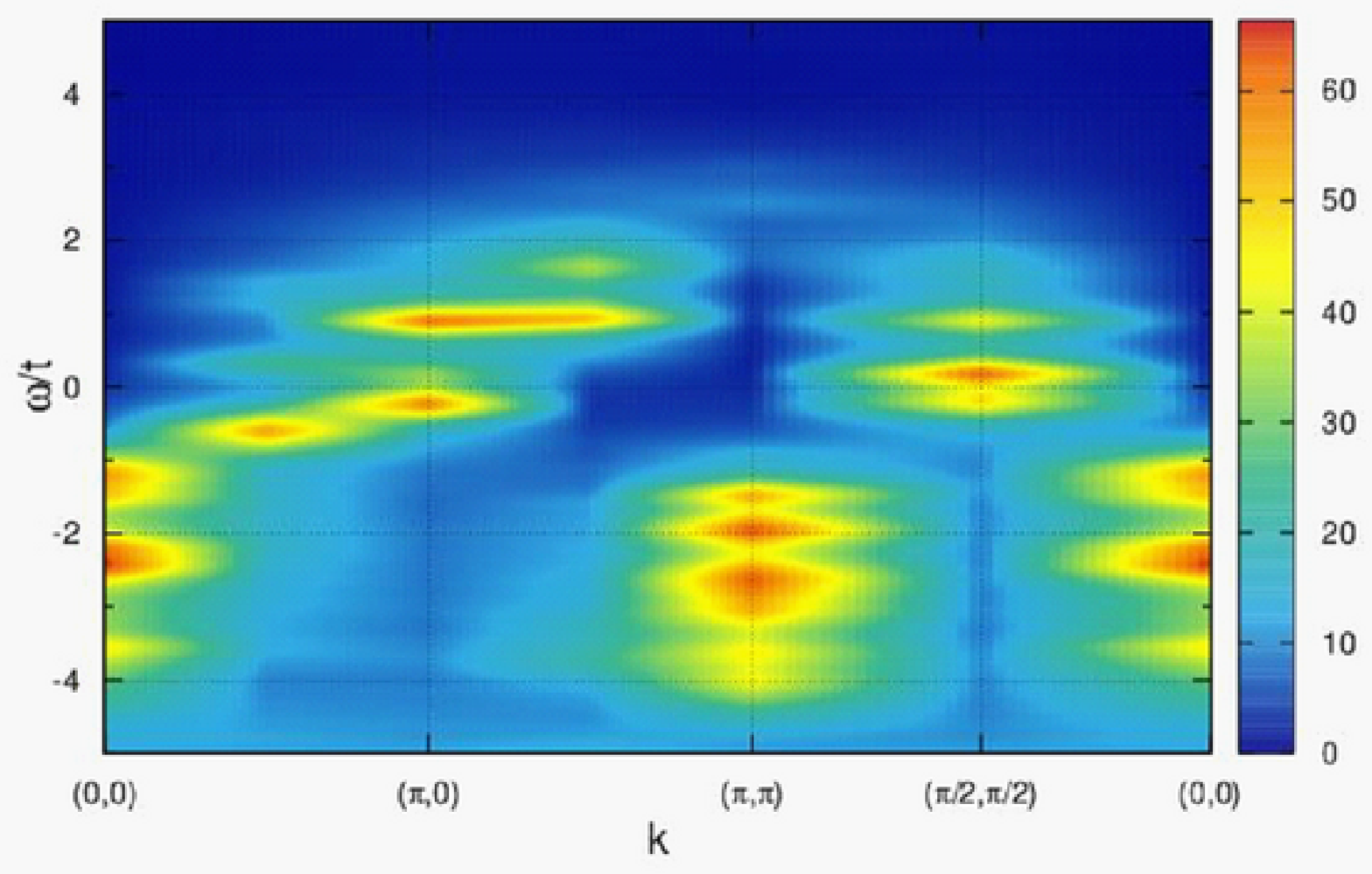}
\end{center}
\caption{Spectrum for $\Psi_{3}^{+}=\vec{n}\cdot\vec{\sigma}\left(\vec{n}\cdot\vec{\sigma}\xi^{\alpha}\right)^{\alpha}$ for the two-hole state}
\label{fig3h}
\end{figure}

Let us move to discuss the band structure in 2D case. We examine the hole doped case ($\delta=0.125$) shown in Figs.~\ref{fig3e}-\ref{fig3h}. For comparison, we also show undoped case in Figs.~\ref{fig3a}-\ref{fig3d}. We should be careful for the fact that the numerical data may involve strong finite-size effects. The excitations $\Psi_{4}^{\mu\nu}$ wrap the overall lattice sites, and their motions are highly restricted. Thus, we only show the spectral data for $\Psi_{1}$, $\Psi_{2}$, $\Psi_{3}^{-}$, and $\Psi_{3}^{+}$. 

We examine the spectrum for $\Psi_{1}=\xi$. The data are shown in Fig.~\ref{fig3e}. In the figure, the Fermi level is located at the origin of the energy $\omega=0$. Comparing this with Refs.~\cite{Tohyama,Kohno}, we see that the result is reasonable. The non-interacting band is highly renormalized, and the band width becomes roughly half. Furthermore, when we look at the spectrum near the Fermi level, we find that the weight at $\vec{k}=(\pi/2,\pi/2)$ and $(\pi,0)$ is very strong. When we look at the branch from $(0,0)$ to $(\pi,\pi)$ through $(\pi,0)$, we find that the weight just above the Fermi level is once reduced at around $(\pi,0)$. This tendency seems to be consistent with the presence of the Fermi arc. We have also confirmed that the gap clearly opens at this momentum by introducing long-range hopping $t^{\prime}$ and $t^{\prime\prime}$. According to Ref.~\cite{Tohyama}, the pseudogap should be also seen just above the Fermi level at around $(\pi/2,\pi/2)$ (This gap is not seen experimentally, since the gap is located at the electron addition side). The branch from $(0,0)$ to $(\pi,\pi)$ through $(\pi/2,\pi/2)$ suffers from the presence of the gap owing to our finite lattice system, and thus unfortunately we cannot determine whether the gap exists. It is also noted that there is a considerable amount of broad spectral weight at around $\vec{k}=(\pi,\pi)$ and $-4\le\omega/t\le -2$.

We also examine the spectra for $\Psi_{2}$, $\Psi_{3}^{-}$, and $\Psi_{3}^{+}$ in Figs.~\ref{fig3f}, \ref{fig3g}, and \ref{fig3h}, respectively. We find that $\Psi_{2}$ and $\Psi_{3}^{+}$ have strong spectral intensity near the Fermi level, while $\Psi_{3}^{-}$ does not have the intensity near the Fermi level. Thus, the strong coupling between the doped hole and the spin cloud is crucial for the intensity even in 2D cases. Furthermore, the intensity near the Fermi level increases as the doped hole is dressed with larger and larger spin cloud. All of the excitations do not have the remarkable spectral intensity above the Fermi level. This feature is also roughly consistent with 1D case, and strongly suggests that the character of the wave function changes at the Fermi level.

\section{Comparison between 1D and 2D results}

It would be quite meaningful to compare the 1D results with the 2D results along $(0,0)$-$(\pi,\pi)$ direction through $k=(\pi/2,\pi/2)$~\cite{Kohno}. For instance, when we look at the spectra for $\Psi_{2}$ in 1D and 2D cases, we see that they are quite similar with each other. This observation suggests that the large spectral intensity at $k\sim(\pi,\pi)$ and $\omega\sim -4t$ in 2D originates in the spinon-like excitation observed in 1D. Originally, if we do not consider the spin fluctuation, there is no spectral weight at around $\vec{k}=(\pi,\pi)$ and $\omega/t\sim -4$. Thus, this is quite reasonable. However, if we look at $\Psi_{3}^{+}$ and $\Psi_{4}^{++}$, the intensity at $\vec{k}\sim(\pi,\pi)$ and $\omega/t\sim -4$ in 2D is much enhanced than that in 1D. This may be due to dimensionality dependence on the coupling strength between spin and charge degrees of freedom. Since the 1D metallic band originates in the holon branch before doping, in 2D case also the weight at around $-4\le\omega/t\le -2$ may be pushed above the Fermi level by the hole doping. This scenario is actually consistent with our conclusion that the character of the wave function changes at the Fermi level.

If the dimensionality dependence is less pronounced on the spectra of composite particles, we may say that the presence of the pseudogap in the 2D $t$-$J$ model and high-$T_{c}$ cuprates can be understood by considering the vanishing spectral weights of the composite excitation above the Fermi level in 1D cases. I believe that this observation is a strong indication that the pseudogap is different from the superconducting gap. However, this is still highly hypothesized conjecture, and thus we need to examine more about the vanishing spectral weight above the Fermi level in 1D and 2D in more physical standpoints.

\section{Summary}

Summarizing, we have examined spectral properties of composite excitations in the 1D and 2D $t$-$J$ models. The most important goal of this numerical work is that these composite hole excitations well characterize prominent band structures in the angle-resolved photoemission spectrum. Therefore, it is possible to do more advanced spectroscopy rather than the standard ARPES spectrum. In 1D case, we have identified that the spinon and holon operators are represented as $\Psi_{2}$, $\Psi_{3}^{+}$, and $\Psi_{4}^{++}$. After doping to the Mott insulator, the spectral weight of $\Psi_{3}^{+}$ and $\Psi_{4}^{++}$ concentrates on the Fermi level, and they are physical excitations. We have also found that the 1D composite data are quite similar to those in 2D except for relatively large spectral intensity of 2D at $\vec{k}\sim(\pi,\pi)$ and $\omega/t\sim -4$. In both of 1D and 2D cases, we have found that the band dispersions above and below the Fermi level have different characters, respectively. We have proposed that this feature would provide us quite important information on the formation of the pseudogap in high-$T_{c}$ cuprates.

\section*{Acknowledgement}

HM acknowledges Masaki Fujita and his co-workers for valuable discussion.

\end{document}